\documentclass[a4paper]{jpconf}
\usepackage{graphicx}

\newcommand{\GeVcc}{\mbox{${\rm GeV/c^2}$}}

\newcommand{\neutt}{$\tilde{\chi}$}
\newcommand{\hetrois}    {\mbox{$ ^{3}{\mathrm{He}\  }                            $}}

\def\JCAP#1#2#3{#2 {\it JCAP} {\bf{#1}}  #3}
\def\PLB#1#2#3{#2  {\it Phys.~Lett.} {\bf{B#1}}  #3}

\begin{document}
\title{MIMAC-\hetrois : A Micro-TPC Matrix of Chambers of \hetrois for direct detection of Wimps}

\author{D. Santos, E. Moulin, F. Mayet, J. Mac\'\i as-P\'erez}

\address{Laboratoire de Physique Subatomique et Cosmologie (CNRS/IN2P3/UJF), \\
 53, Av. des Martyrs, 38026 Grenoble, FRANCE}

\ead{Daniel.Santos@lpsc.in2p3.fr}

\begin{abstract}
The project of a micro-TPC matrix of chambers of \hetrois 
for direct detection of non-baryonic dark matter is 
presented. The privileged properties of \hetrois  are highlighted. The double detection (ionization - projection of tracks) is explained and its 
rejection evaluated. 
The potentialities of MIMAC-\hetrois   
for supersymmetric dark matter search 
are discussed 
\end{abstract}

\section{Introduction}

In the last years our work on \hetrois as a target for detecting WIMPs  allowed to confirm its 
privileged properties for direct detection \cite{idm2002}. These properties can be enumerated  as follow : 		
\begin{itemize}
\item  its fermionic character opens the axial interaction with fermionic WIMPs as the neutralinos,
\item the extremely low Compton cross section reduces by several order of 
magnitude the natural radioactive background with respect to other targets,
\item the high neutron capture cross section gives a clear signature for neutron rejection,
\item its light mass allows a higher sensitivity to light WIMP masses than other targets, 
\item the elastic energy transfer is bounded to a very narrow range of energy (a few keV) offering a high signal to noise ratio.
\end{itemize}

The extremely low Compton cross section and the possibility to detect events in the keV range 
($\leq $5.6 keV) have been demonstrated by the $^{57}$Co electron conversion 
detection recently reported \cite{electrons}. The detection of  $\rm 7 \ keV$ electrons 
in the MACHe3 prototype with the source emitting  
121 keV $\gamma$-rays embedded in the \hetrois is a clear demonstration 
of the virtual transparency of this medium to the electromagnetic radiation.

\noindent
\section{Micro-TPC and ionization-track projection}

\begin{figure}[t]
\begin{center}
\includegraphics[scale=0.5]{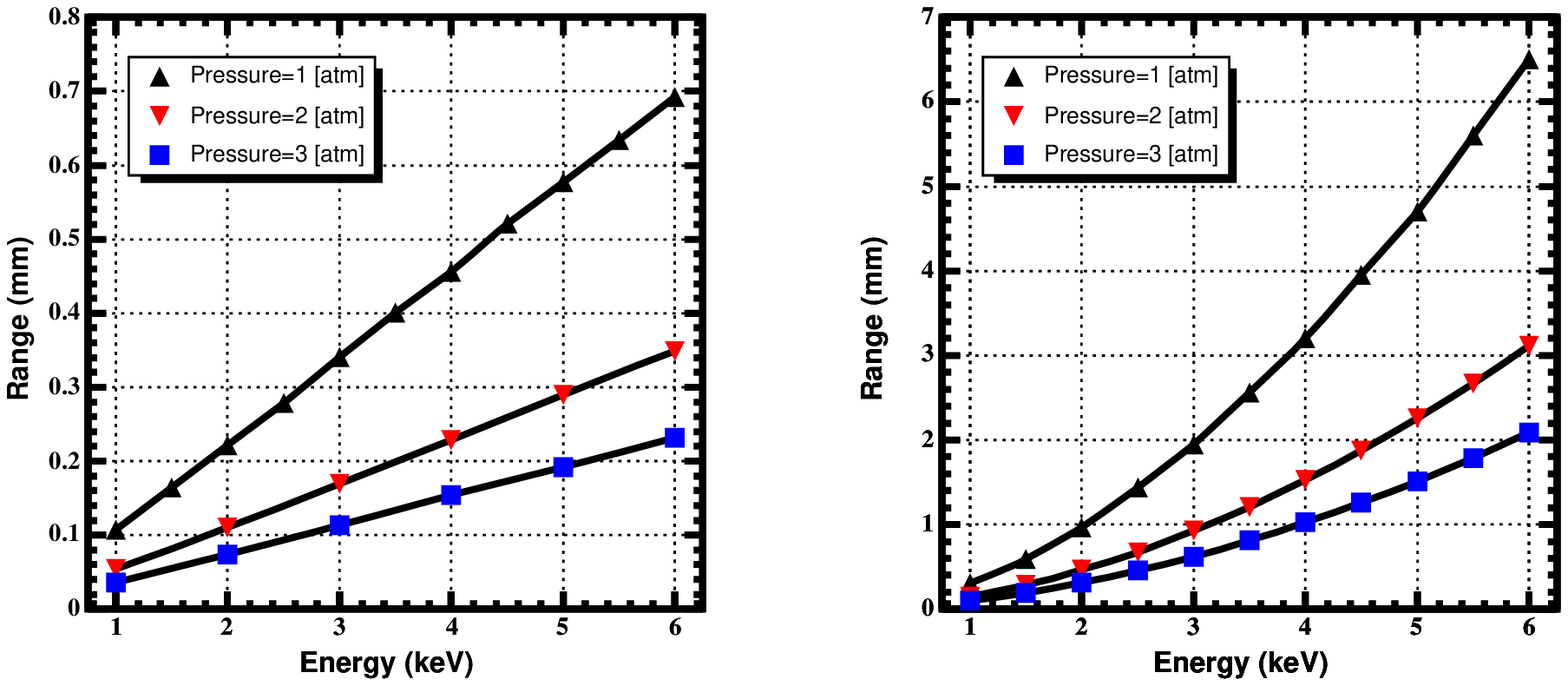}
\includegraphics[scale=0.5]{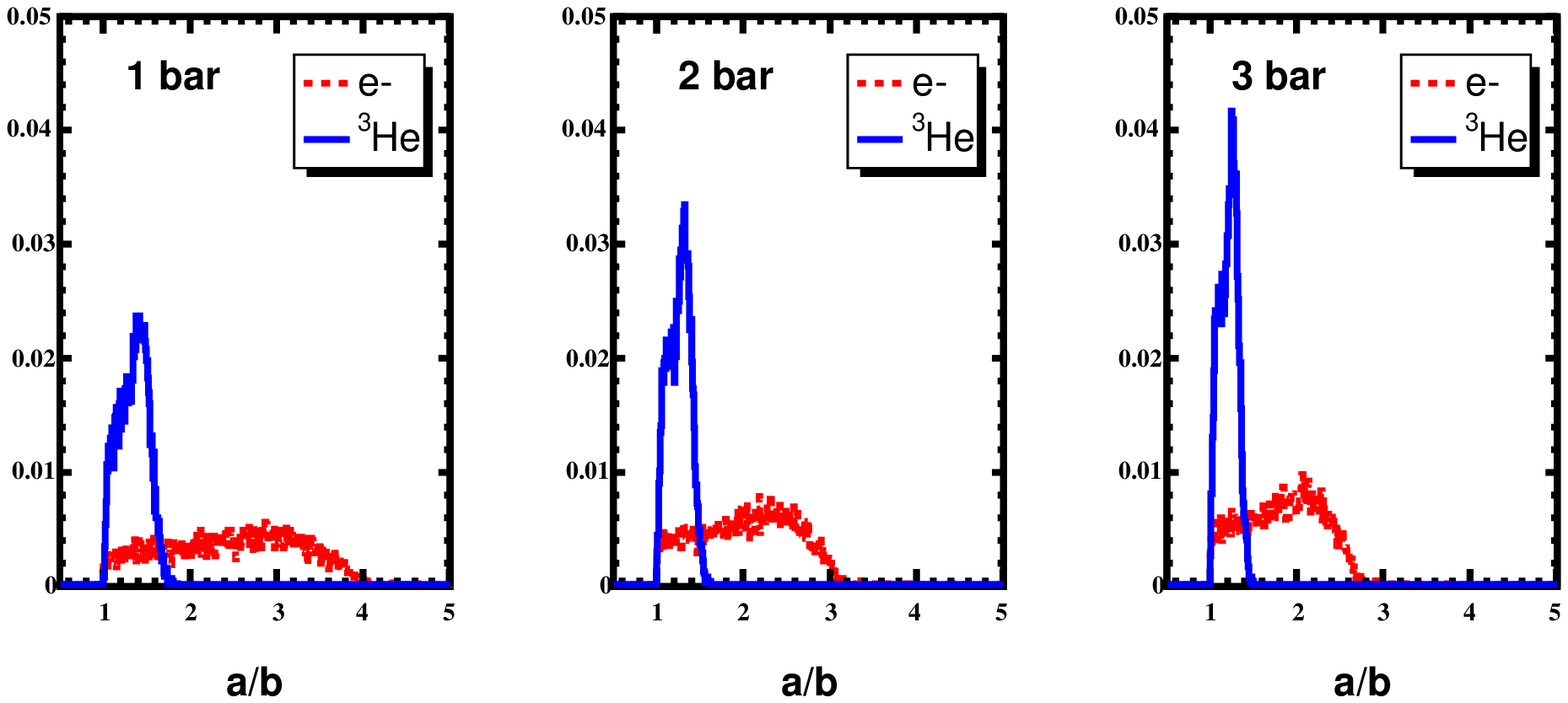}
\caption{\label{range} Upper panel : Simulated 
range vs kinetic energy for  \hetrois  (left) and electrons (right), at 3 pressures. Lower panel : 
Simulated distribution  of the ratio a/b  at 
different pressures.}
\end{center}
\end{figure}

The micro temporal projection chambers with an avalanche amplification using a pixelized  
anode presents the required features to discriminate electron - recoil events with the 
double detection of the ionization energy and the track projection onto the anode.  
In order to get the electron-recoil discrimination, the pressure of the TPC should be 
such that the electron tracks with an energy less than 6 keV could be well resolved from 
the recoil ones at the same energy convoluted by the quenching factor. 
Simulations have been done, as a function of the pressure,  
for electrons using Geant 4 and for recoils using SRIM. The results are shown on fig. \ref{range}. 

The electrons produced by the primary interactions will drift to the grid in a diffusion process following 
the well known distribution characterized by a radius of $\rm D\simeq 200\mu m \sqrt(L[cm]) $ where $\rm L$ 
is the total drift in the chamber up to the grid. This process has been simulated with Garfield 
and the drift velocities estimated  as a function of the pressure and the electric field. 
A typical value of $\rm 26 \mu$m/ns is obtained for $\rm 1\ kV/cm$ at a pressure of 1 bar. 
To prevent confusion between electron track projection and recoil ones the 
total drift length should be limited to L$\simeq $15 cm. It defines the 
elementary cell of the detector matrix and the simulations performed on the ranges of electrons 
and recoils suggest that with an anode of $\rm 350 \mu m$  the electron-recoil discrimination required can be obtained.
The quenching factor is an important point that should be addressed to quantify the 
amount of the total recoil energy recovered in the ionization channel. No measurements of the 
quenching factor (QF) in \hetrois have been  reported. However, an estimation can be obtained applying   the Lindhard calculations
\cite{lindhard}.  
The estimated quenching factor given by SRIM for \hetrois shows up to 70 \% of the recoil energy 
going to the ionization channel for 5 keV  \hetrois recoil. 

To measure the QF in \hetrois  at such low energies, 
we have designed, at the LPSC laboratory, an ion source to accelerate 
the helium ions  before entering  to the micro-TPC chamber. 
They will pass through a
 thin foil of polypropylene that will neutralize them. The measurement of the energy of the atoms of \hetrois 
  will be made by a time of flight measurement.

\noindent
In order to characterize the distribution of pixels on the anode for various  
trajectories we define the ratio between perpendicular 
symmetry axis of the pixel distribution (a/b) where a is the larger axis of the distribution.
We plot on fig.~\ref{range}   the simulated distribution  of the ratio a/b  at 
different pressures. An isotropic spherical emission of electrons and recoils at $\rm L \sim 10 \ cm$ 
from the grid has been injected as the input of the simulation. For the recoils a very 
concentrated distribution around 1 is expected, and for the electrons a very wide  one.
The rejection of events using the a/b ratio is a 
strong function of the energy and the pressure of the chamber, 
but even at 1 keV and 3 bar only a small number  of  the total events can be confused.

\section{Supersymmetric dark matter search}
\begin{figure}[th]
\begin{center}
\mbox{\hspace{-.5cm}\includegraphics[scale=0.34]{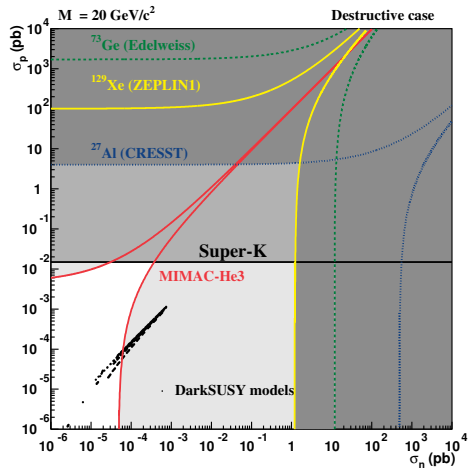}}
\caption{\label{fig:sigpsigndirind} SUSY models satisfying accelerator and 
cosmological constraints (black points) in the SD cross-section on proton  versus   neutron plane 
for 20 \GeVcc \  neutralino.}
\end{center}
\end{figure}

The potentialities of MIMAC-\hetrois   
for supersymmetric dark matter search 
are discussed \cite{moulinplb} within the framework of effective MSSM models
without gaugino mass unification at the GUT scale.  
Indeed, the unification constraint can be relaxed with a free parameter R defined by  
${\rm M_1\,\equiv\,R\,M_2}$. In this case, the neutralino \neutt \ can be lighter 
than the LEP limit obtained in universal scenarii. 
A large scan of the parameter space has been done with the DarkSUSY code~\cite{ds} in which the departure from universality 
(${\rm R\,<\,0.5}$)  has been implemented. The complementarity between various detection strategies for spin dependent (SD)
interaction have been studied. 
Figure~\ref{fig:sigpsigndirind} shows the result in the (${\rm \sigma^{SD}_p, \sigma^{SD}_n}$) plane 
in the destructive   interference case, 
for 20 \GeVcc \  neutralinos. Exclusion curves from CRESST, ZEPLIN-I and 
Edelweiss are presented (see \cite{moulinplb} for details). 
For a given exclusion limit, the excluded region lies outside the two curves.

\noindent
The current excluded region (dark and medium grey) in this plane is given by the 
combination of these curves. It also includes the limit from indirect DM detection, from Super-K limit\ref{fig:sigpsigndirind}, which is   
exclusively sensitive to the SD cross-section on proton. Therefore, this constraint 
strongly reduces the allowed region with a near orthogonality to neutron based experiments, whereas 
proton based experiments (CRESST) are well overlapped by Super-K limit. 
However, SUSY models (black points) neither excluded by accelerator  nor 
cosmological constraints lie well below this limit.  
The projected exclusion curve for MIMAC-\hetrois is displayed. It can be seen that most of 20 \GeVcc \ 
neutralinos, escaping from detection of ongoing experiments, would be visible by MIMAC-\hetrois. 
This study  highlights the complementarity of this experiment with 
most of current spin-independent and spin-dependent  experiments : proton based detectors as well as $\nu$ telescopes.

\smallskip
 
\end{document}